\documentclass[twocolumn]{aastex62}

\received{}
\revised{}
\submitjournal{AJ}

\shorttitle{A Multi-Band Star Cluster Catalog}
\shortauthors{Bica et al.}

\begin{document}

\title{A Multi-Band Catalog of 10978 Star Clusters, Associations and Candidates in the Milky Way}

\correspondingauthor{Charles J. Bonatto}
\email{charles.bonatto@ufrgs.br}

\author{Eduardo Bica}
\affiliation{Departamento de Astronomia \\
Universidade Federal do Rio Grande do Sul \\
Av. Bento Gon\c calves 9500\ \\
Porto Alegre 91501-970, RS, Brazil}

\author{Daniela B. Pavani}
\affiliation{Departamento de Astronomia \\
Universidade Federal do Rio Grande do Sul \\
Av. Bento Gon\c calves 9500\ \\
Porto Alegre 91501-970, RS, Brazil}

\author{Charles J. Bonatto}
\affiliation{Departamento de Astronomia \\
Universidade Federal do Rio Grande do Sul \\
Av. Bento Gon\c calves 9500\ \\
Porto Alegre 91501-970, RS, Brazil}

\author{Eliade F. Lima}
\affiliation{Universidade Federal do Pampa \\
Br 472 - Km 585, CP\,118 \\
Uruguaiana, RS, Brazil}



\begin{abstract}

We present a catalog of Galactic star clusters, associations and candidates with 10978 entries. 
This multi-band catalog was constructed over 20 years, starting with visual inspections on the 
Digital Sky Survey and incremented with the 2MASS, WISE, VVV, Spitzer and Herschel surveys. Large 
and small catalogs, 
as well as papers on individual objects have been  systematically cross-identified. The catalog 
provides Galactic and equatorial coordinates, angular diameters, and chronologically ordered 
designations, making it simple to assign discoveries, and verify how often the objects were 
cataloged by different authors, search methods and/or surveys. Detection in a single band is 
the minimum constraint to validate an entry. About 3200 objects have measured  parameters in  
the literature. A fundamental contribution of the present study is to present additional 
$\approx7700$ objects for first analyses of nature, photometry, spectroscopy and structure. 
The present focus is not to compile or determine fundamental parameters, but to provide a catalog 
characterizing uniformly the entries. A major result is that now 4060 embedded clusters are 
catologed, a factor of $\approx2$ larger than open clusters. Besides cross-identifications in 
different references and wavelength domains, we also communicate the discovery of 638 star 
clusters and alike objects. The present general catalog  provides previously studied objects 
and thousands of additional entries in a homogeneous way, a timely contribution to {\em Gaia} related
works.

\end{abstract}

\keywords{astronomical databases: atlases, catalogs, surveys --- Galaxy: bulge, disk, halo,
general --- (Galaxy:) globular clusters --- Galaxy: open clusters and associations: general }


\section{Introduction} 
\label{sec:1}

The first  large  catalog of star clusters compiled in the  literature  including discoveries  was that of  
\citet{Collinder1931}, with 471 open clusters (OCs). \citet{Alter1970} and \citet{Lynga1987}  gathered nearly  
1000 OCs each. Two databases were developed by collecting OCs and their fundamental parameters: WEBDA 
(\citealt{Netopil2012}) and DAML02 \citep{Dias2002}, with $\approx1200$  and 2167  clusters, respectively. 
After these first discoveries and taking the importance of investigation of star clusters for the 
formation and evolution of the Galaxy into account, as well as its kinematics, the studies in the field 
are rapidly growing. 

More recently, \citet{Kharchenkoetal2013} employed  a stellar base including 2MASS,  and homogeneous  analysis 
tools  to obtain parameters  for  3006  star clusters and associations. It would be important now to create an 
overall catalog of star clusters and alike objects (SCAOs) in the Milky Way to homogeneously give  previous  
objects  and new entries for photometric and structural analyses. Such catalog should encompass OCs, globular 
(GC) and embedded clusters (ECs), their candidates and alike classes, such  as  associations and different types  
of stellar groups. The search of depopulated open cluster  remnants  (e.g. \citealt{PavaniBica2007}) may lead 
to evolved dynamical stages. This  together with  low mass dissolving ECs \citep{Oliveiraetal2018} are important 
to understand how stars feed the Galactic field (\citealt{Lada2003}). The present catalog (hereafter CatClu) 
cross-identifies all previous SCAOs and presents newly found ones. 

As a byproduct of the present catalog  we give in the appendix CatKGr, a compilation of stellar groups that did 
not fit CatClu. Part of them are kinematic groups such as halo streams, e.g. \citet{BalbGi2018}. An additional  
byproduct is CatGal given also in the appendix. It is an updated list of the Local Group galaxies and references 
to their star clusters and associations which is a fast developing field. CatGal includes ultrafaint dwarf 
galaxies (e.g. \citealt{Drlica-Wagneretal2015}), some of which were revised  to  faint halo star clusters, 
becoming a source of new objects for CatClu. CatClu includes a subcatalog of 640 hereby found clusters 
and candidates. In Sect.~\ref{sec:2} we describe the data sources. Sect.~\ref{sec:3} describes  how the 
database was constructed. In Sect.~\ref{sec:4}, the angular distributions of different object types is discussed.  
Finally, in Sect.~\ref{sec:5} the concluding remarks are given.

\section{Data Sources}
\label{sec:2}

In this work, both large (typically more than 1000 entries) and small catalogs were analysed. Also, small  sample 
papers were taken into account. Catalogs are in general related to specific surveys. Examples are \citet{Bicaetal2003a}, 
\citet{Dutraetal2003} and \citet{Froebrichetal2007} with 2MASS, \citet{Merceretal2005} with Spitzer, and  
\citet{Solinetal2012} with UKIDSS. Concerning  the ESO-VISTA VVV survey, \citet{Borissovaetal2011}, \citet{Borissovaetal2014} 
and \citet{Barba2015} provided new clusters towards the bulge and central disk, which are as a rule  absorbed ones. 
\citet{Limaetal2014} found new clusters in the NGC\,6357 complex with VVV. \citet{Majaess2013}, \citet{CBB2015a} and 
\citet{CBB2016a} employed  WISE in their EC discoveries. In the present study, the Aladin Sky Atlas  with several 
surveys therein was used to cross-identify catalogs and individual cluster studies. The main observational 
aspects for classifying objects are related to morphology. We analyse the central concentration, hierarchical
structures, stellar density of the object in contrast to the field, Red, Green and Blue band combinations for 
stellar colors, and the presence of dust and gas emission. Finally, we estimate central coordinates and angular 
dimensions. Previous analyses in the literature are also considered.

For an overview of this work, we anticipate in Table~\ref{tab1} the statistics of the derived catalogs in this 
study (Sect.~\ref{sec:3} and appendix). Thus, the paper contents can be appreciated by means of the object  
classifications and their counts.  By columns: (1) object class abbreviation, (2)  object class, (3) catalog 
table, (4) population counts.


\begin{deluxetable*}{llcr}
\tablecaption{Object Classes and their Counts\label{tab1}}
\tablehead{\colhead{Code}&\colhead{Object Classification}&\colhead{Table}&\colhead{Entries}}
\colnumbers
\startdata
OC                 & Open Clusters                           & 3   & 2912\\
OCC                & Open Cluster Candidates                 & 3   &  651\\
\hline  
OC $+$ OCC         & Sum of OPEN CLUSTERS \& ALIKE           & 3   & 3563\\
\hline  
EC                 & Embedded Clusters                       & 3   & 4234\\
ECC                & Embedded Cluster Candidates             & 3   &  349\\
EGr                & Embedded Groups                         & 3   &  354\\
\hline  
EC $+$ ECC $+$ EGr & Sum of EMBEDDED CLUSTERS \& ALIKE       & 3   & 4937\\     
\hline           
lPOCR              & loose Open Cluster Remnant Candidates   & 3   &  449\\
cPOCR              & compact Open Cluster Remnant Candidates & 3   &   78\\
\hline  
lPOCR $+$ cPOCR    & Sum of OPEN CLUSTER REMNANT CANDIDATES  & 3   &  527\\
\hline  
Assoc              & Associations                            & 3   &  470\\
GC                 & Globular Clusters                       & 3   &  200\\
GCC                & Globular Cluster Candidates             & 3   &   94\\
\hline  
GC + GCC           & Sum of GLOBULAR CLUSTERS AND CANDIDATES & 3   &  294\\
\hline  
MHC                & Magellanic Halos's Clusters             & 3  &   33\\
Ast                & Asterisms                               & 3   & 1154\\                                                                
\hline
&      \multicolumn{2}{c}{Total of individual entries}              &10978\\
\hline
KGr                & Kinematical Groups                      & 4   & 228 \\
KAs                & Kinematical Associations                & 4   & 17 \\
KGr + KAs          & Sum of KINEMATICAL GROUPS AND ASSOCIATIONS & 4   & 245 \\     
DGAL               & Local Group Dwarf Galaxies               & 5   & 138 \\
NGAL               & Local Group Normal Galaxies              & 5   & 6   \\
DGAL + NGAL        & Sum of LOCAL GROUP GALAXIES              & 5   & 144 \\
\hline  
&\multicolumn{2}{c}{Total entries in the database}& 11367\\
\hline  
\enddata
\tablecomments{Candidates correspond to the codes OCC, GCC and ECC. The EGr class  
corresponds to looser and less populated entries than ECs \citep{Bicaetal2003b}. }
\end{deluxetable*}

Electronic Table~\ref{tab2} provides the references for the 3 catalogs in the present work (Table~\ref{tab3}, 
Table~\ref{tab4} and Table~\ref{tab5}). By columns: (1) object class, (2) number of relevant objects extracted 
from the reference, (3) designation or acronyms, (4) reference code, (5) bibliographic reference. Table~\ref{tab2} 
includes 792 references,  the last 14 are electronic. As examples we comment the first 5 entries. Herbig\,1 and 
Manova\,1 are two previously overlooked clusters retrieved here  as ECs,  owing to their related dust emission in 
WISE. Herbig\,1 has currently a more recent designation  in SIMBAD. Manova\,1 is a new entry not present in SIMBAD. 
This emphasizes the historical and  chronological search that we made throughout essentially all the literature.  
$\epsilon$ Indi, 61\,Cyg, Gamma Leo and Groombridge\,1830 are stellar groups named after a representative  member. 
Owing to proximity they are defined not by their coordinates, but by their heliocentric U, V and W velocities 
(e.g. \citealt{Eggen58}). They are stellar groups that do not fit star cluster characterizations. Together with 
moving groups and streams they are classified as the general term kinematic group (Table~\ref{sec:4}).

\begin{deluxetable*}{lllll}
\tablecaption{References for the CatClu, CatKGr and CatGal Catalogs\label{tab2}}
\tablehead{\colhead{Class}&\colhead{N}&\colhead{Name/Acronym}&\colhead{Code}&\colhead{Reference}}
\colnumbers
\startdata
EC     & 1 & Herbig\,1          & 103  &\citet{Herbig58} \\
KGr    & 2 & epsilon Indi, 61\,Cyg & 1743 & \citet{Eggen58} \\
KGr    & 1 & Gamma Leo          & 1737 &\citet{Eggen59} \\
KGr    & 1 & Groombridge\,1830  & 1738 &\citet{EggenSandage59} \\
EC     & 1 & Manova\,1          & 3044 & \citet{Manova59} \\
\enddata
\end{deluxetable*}


The references in Table~\ref{tab2} include more than 30000 items that were analysed one by one over 20 years 
to infer or verify their nature, characterization and cross-identification. We emphasize that these procedures 
were carried out  essentially independent of SIMBAD. Both archiving approaches in general provide similar object 
data, except that chronology is not systematic in SIMBAD.

We conclude that Table~\ref{tab2} is a tool in itself with 792 references to inject information into and from 
the CatClu (Table~\ref{tab3}), CatKGr (Table~\ref{tab4}) and CatGal (Table~\ref{sec:5}) catalogs. This large 
reference set is provided in the electronic table format, and is thus not part of the paper itself references. 
In the following we discuss classifications.

\subsection{Embedded Clusters, Open Clusters and Associations}
\label{sec:2.1}

We use \citet{Lada2003} classification of ECs and OCs. Earlier studies referred to ECs as OCs within  nebula 
and/or dust. \citet{Lada2003} provided a physical classification taking gas and dust loss in the cluster into 
account. Concerning morphology and structure of ECs, they distinguish centrally condensed or
hierarchical clusters. \citet{Ascenso18} introduces additional criteria, such as the presence of sub-structures, 
multiple nuclei, and fractal distribution. All these criteria were applied to the available observational images.
In future studies, it would be interesting to break up our EC classification into such sub-types.

Mass-loss processes in the early evolution of ECs as a rule dissolves them, \citep{T1978}. In this scenario OCs 
are $\approx5\%$  of the ECs, which  dynamically survive the gravitational potential loss. Recognizing ECs  is 
straightforward by their connection  with gas and dust, especially in the IR domain showing dust emission in WISE, 
Spitzer and  Herschel. We suggest the inclusion  of the EC classification in SIMBAD. OB associations are in general 
extended structures with  massive stars that are looser than star clusters and occur along  spiral arms 
\citep{Ambartsumian1955}. R associations are groups of reflection nebulae, themselves sites of star formation 
\citep{HR1976}. T associations are star-forming stellar groups in a stage  that contains  many T Tauri stars 
\citep{Ambartsumian1957}. Associations sometimes present subassociations \citep{Mel1995}. HIPPARCOS organized 
nearby associations by means of positions, proper motions (PMs) and parallaxes \citep{deZeeuwetal1999}, while 
{\em Gaia} will have a fundamental role in the  definition of associations beyond our neighborhood. The present 
study collected and cross-identified, and condensed them into 470 associations (Table~\ref{tab1}), which are 
provided in Table~\ref{tab3} with the respective references (Table~\ref{tab2}). Since  massive ECs  can also 
become unstable and dissolve, they may   contribute to  the development of  associations (\citealt{Saurin2012} 
and references therein).

\subsection{Cluster Remnants and Candidates}
\label{sec:2.2}

Poorly populated stellar concentrations can turn out to be OC remnants, or field fluctuations, when studied 
by means of color-magnitude diagrams (CMDs) and other means (e.g. \citealt{Carraro2000}, \citealt{Ode2002}, 
\citealt{PavaniBica2007}, \citealt{Pavanietal2011}). We have suggested two criteria to observationally define 
the candidates: (i) poorly populated stellar concentrations, and/or (ii) stellar surface density variations along 
the cluster position angle on the sky \citep{Bicaetal2001}. The loose Possible Open Cluster Remnants (lPOCR) are 
very common (Table~\ref{tab1}). We also provide an updated list of compact Possible Open Cluster Remnants 
(cPOCR) which are rarer \citep{Pavanietal2011}. The former are expected to be dynamically evolved OCs, and the 
latter fossil cluster cores \citep{Bicaetal2001}. Recently, some dynamically advanced OCs have been proposed by 
\citet{Angelo18} and further support evolutionary connections.

\subsection{Asterisms }
\label{sec:2.3}

Asterisms are stellar configurations or concentrations that are not expected to be star clusters or associations. 
Many have been identified by amateur astronomers, often with telescopes or in DSS. Under close scrutiny, some 
of them turned out to be star clusters (\citealt{BicaBonatto2011}, \citealt{Dias2002}). Their classification also 
encompasses poorly populated stellar concentrations that the literature showed not to be star clusters (e.g. 
\citealt{Ode2002}, \citealt{Pavanietal2011}). Data-mining the asterisms in Table~\ref{tab3} and analyzing them  
with photometry and PMs, in particular with {\em Gaia}, will certainly reveal a number of new interesting clusters. 
The amateur astronomer group Deep Sky Hunters (DSH) searched for star clusters especially on DSS images and provided 
many candidates that were confirmed to be clusters (\citealt{Kronbergeretal2006}, \citealt{BonBi2010}). Table~\ref{tab1} 
indicates 1154 asterisms which are listed in Table~\ref{tab3}. Several lists are available in the WEB, and others 
were deactivated. Such lists were made by amateur astronomers. One of us (E.B) collected them during 
$\approx1$ decade preserving the original designations, and inspected images with Aladin. Density contrast, 
richness, and relation (or not) to gas and dust were taken into account (Table~\ref{tab3}). Table~\ref{tab2} 
shows 14 electronic references mostly concerning asterisms. An example of a currently active list in the WEB 
is Ferrero's with 53 entries. The largest list was compiled by B. Alessi with 1070 entries as given in DAMLO2.

\subsection{Designations}
\label{sec:2.4}

Historical designations like NGC and IC are easy to remember. In the 20th century an author last name was usually 
employed for OCs (e.g. Trumpler, Markarian or Melotte), or the institute (Harvard), see \citet{Collinder1931}.  In 
the 70's the IAU and C acronyms followed by B1950.0 equatorial coordinates were proposed for star clusters, but  
became obsolete with J2000.0.  Acronyms formed by the author's last name initial letter(s) are usual since the last  
decade of the 20th century (e.g. FSR \citealt{Froebrichetal2007}). Recently,  the acronym MWSC (Milky Way Star Cluster) 
was employed by \citet{Kharchenkoetal2013}. Technical designations at times given  in SIMBAD are complex, with author(s) 
last name initial letter(s)  and the year within brackets, followed by an ordering number, J2000.0 equatorial,  or 
Galactic coordinates. They are an important option for archiving purposes because they are unique among all classes of 
astronomical objects. A disadvantage is their complexity which may lead either to usage simplifications or alternatively,  
doom an object to oblivion. The present catalog adopts all designations which are commonly used  in papers, and are unique 
within the star cluster and  association areas. Additionally, some options must be established. \citet{Dias2002} adopted 
up to two author last  names.  We rather adopt a single author name, and initial last name  letters for 2 or more authors 
(see the case of AL\,2, the second entry  in  Table~\ref{tab3}).  Recently,  \citet{Minniti2017a}  employed the first author 
name as acronym. This is consistent, if a given author searched himself  for new objects  in a survey. In their recent new 
sample  of 84 GC  candidates  they also suggested a shorter  acronym  (Minni),  for simplicity  in papers. Accordingly, we 
suggest the use of the acronym Bica for the objects hereby discovered (Sect.~\ref{sec:3.2}). We recall that the OCs Bica\,1 
through 6 were so designated by \citet{Dias2002}.  We  suggest Bc, for short.

The VVV survey has produced numerous new clusters (e.g. \citealt{Borissovaetal2011}; \citealt{Borissovaetal2014}; 
\citealt{Barba2015}) with designations in terms of VVV-CL and La Serena (LS). With the advent of the VVVX survey 
surrounding that of VVV, discoveries can be designated as VVVX-CL or VVVX for conciseness. In the case of VVVX,
the present catalog will be a fundamental tool for unambiguous new discoveries. 

\section{Catalog Construction}
\label{sec:3}

As a previous experience, \citet{Bica1995} and  \citet{Bicaetal1999}  provided deep catalogs of the LMC and 
SMC-Bridge SCAOs, with 6659 and 1188 entries, respectively. They made cross-identifications using  ESO/SERC 
Red and J (Blue) films. \citet{Bicaetal2008} updated them to 9305 entries. The acquired experience by one of 
us (E.B.) in the  Clouds was fundamental to  collect and analyze SCAOs counterparts in the Galaxy. Together 
with  the  clusters and candidates  in the literature, new ones  were  systematically cross-identified and 
eventually became discoveries.
       
\subsection{A Multi-Band Catalog}
\label{sec:3.1}       
       
Wavelength ranges of the surveys  and their Red, Green and Blue (RGB) color compositions  became wider  with time.  Optical and 
IR survey bands became  available in the  Aladin and IPAC tools in  subsequent upgraded   versions.  This multi-band study  
expanded  from the early DSS to the near-IR  employing 2MASS and VVV, and  further to the mid and far IR with WISE,  Spitzer 
and Herschel.

The approach follows previous catalogs and cluster analyses. Examples are the cluster candidates probed with 2MASS towards 
the bulge and central disk \citep{DutraBica2000}, together with one of  the first cluster general catalogs in the near IR 
\citep{Bicaetal2003b}, and samples  for  detailed studies of CMDs and structure (e.g.  \citealt{BonBi2009}). 

To date,  taking into account WEBDA, DAMLO2 and MWSC (\citealt{Kharchenkoetal2013}, \citealt{Schmejaetal2014}) about 3.200 objects   
have astrophysical parameters. We analysed them together with several thousand new ones in a homogeneous way.  We note that a single 
cluster identification in a given band is enough to be included in the catalog. This stems from the properties of a cluster 
and its line of sight, such as absorption, richness, age and distance, as well as observational 
and instrumental conditions such as seeing, pixel size, crowding, and limiting magnitude of a 
given survey.

\begin{figure}
\plotone{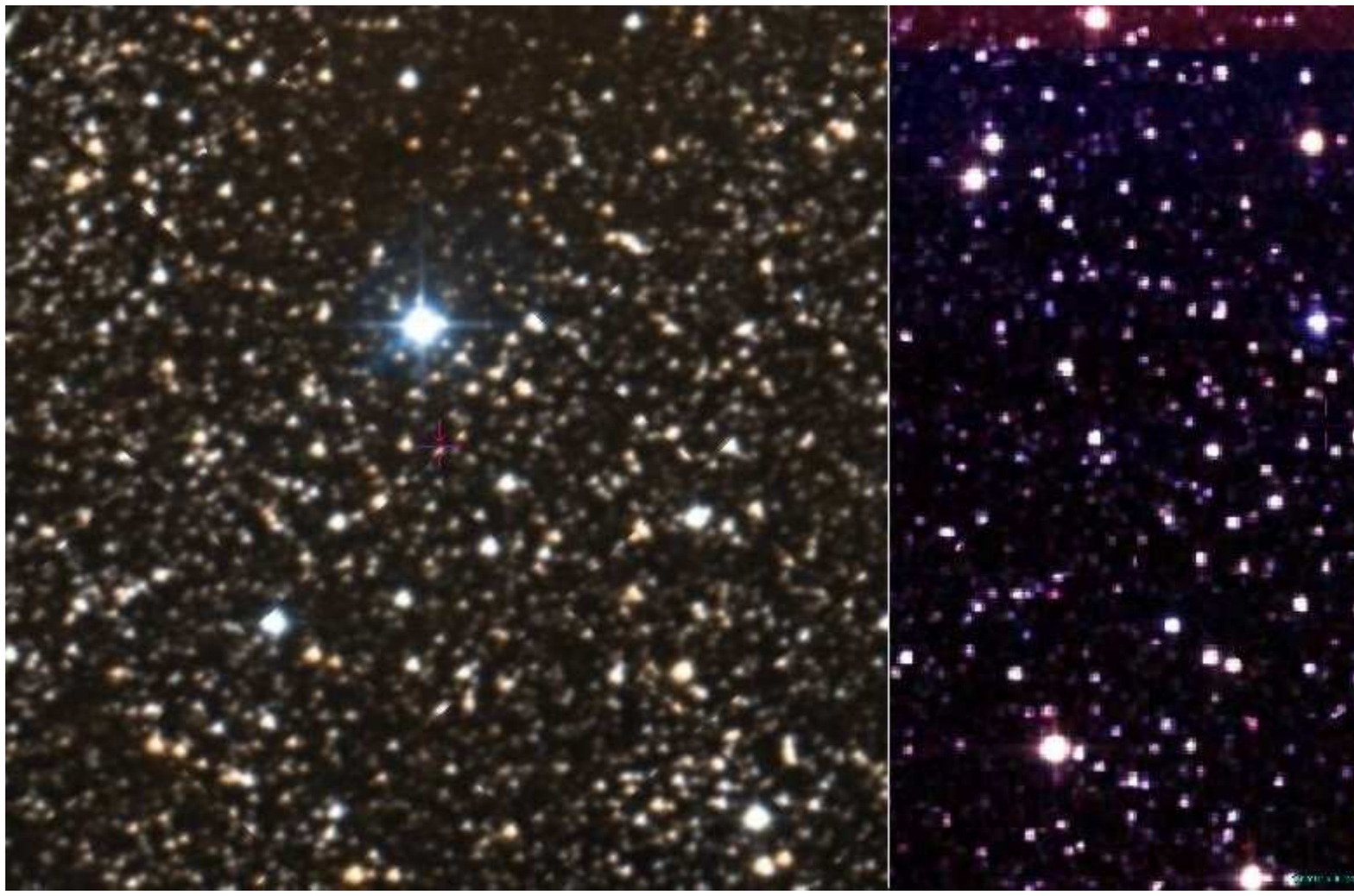}
\caption{Color extractions of $8'\times8'$ for the OCC Bc\,105. North to the top, East to the left.  
Mosaic images from left to right are from DSS, 2MASS and WISE.}
\label{f1}
\end{figure}

\begin{figure}
\plotone{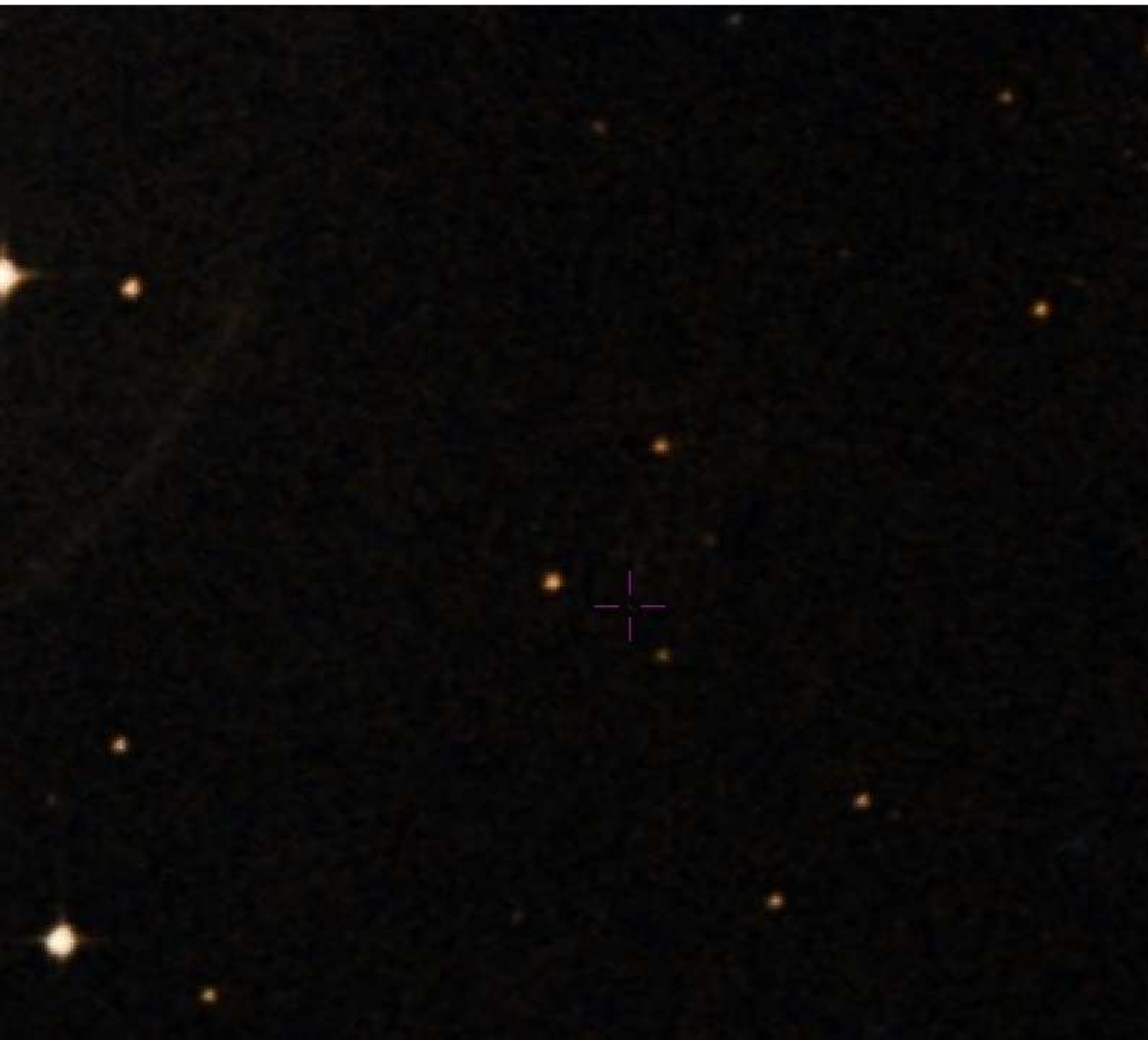}
\caption{Color extractions for the EC Bc\,74 corresponding to DSS, 2MASS, WISE and  Herschel. Angular 
dimensions and orientation as in Fig.~\ref{f1}.}
\label{f2}
\end{figure}

\begin{figure}
\plotone{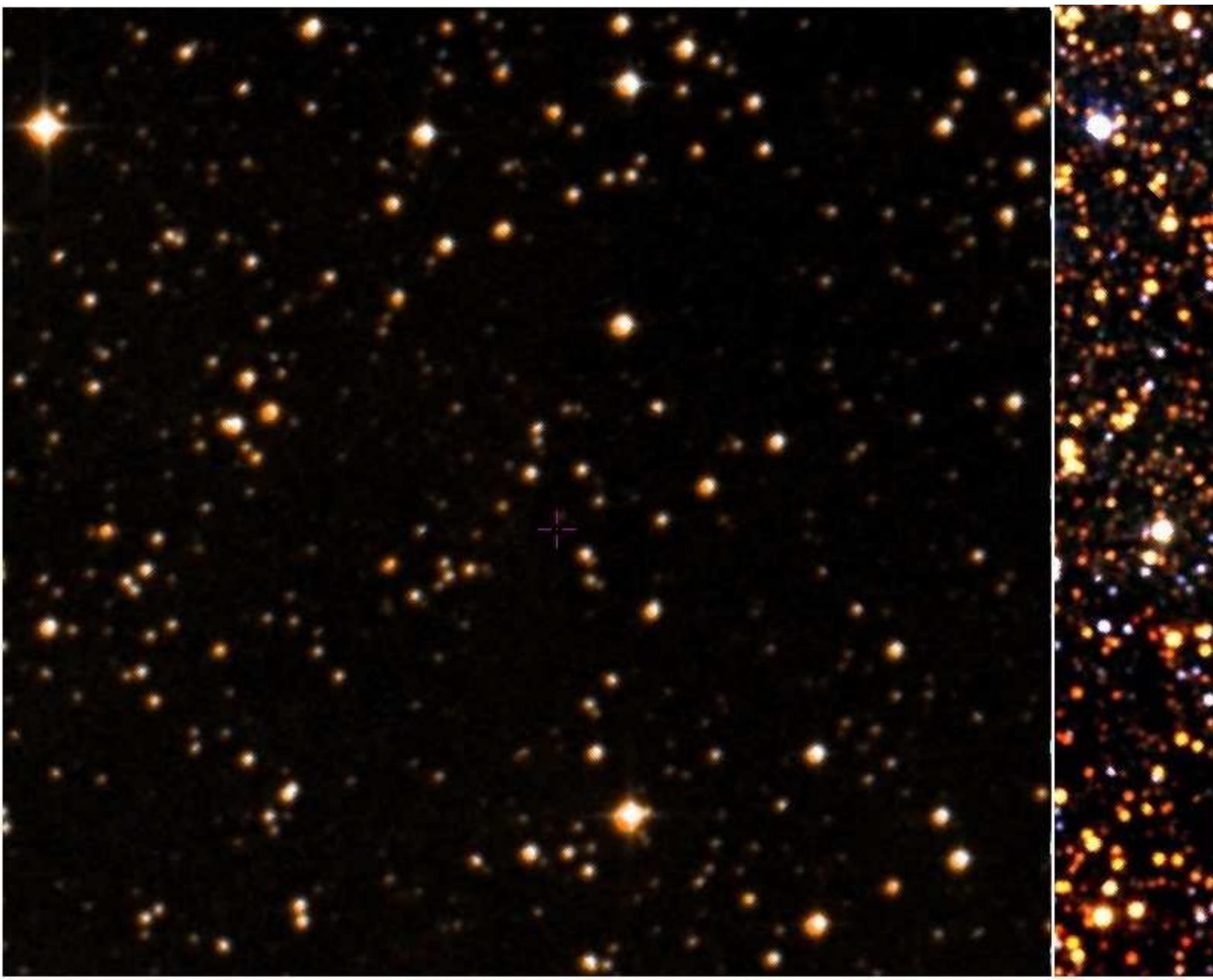}
\caption{Same as Fig.~2, except the third panel which is from Spitzer, for the deeply embedded cluster  Bc\,37.}
\label{f3}
\end{figure}

To illustrate  the meaning  of the multi-band analysis we give  in Fig.~\ref{f1} to  Fig.~\ref{f3} examples of band combinations 
into  RGB colors for three newly found objects (Sect.\,\ref{sec:3.2}). The images of the  surveys  were obtained in Aladin 
version 9.0.   The DSS color atlas (hereafter CA) uses the B, R and I bands.  The 2MASS CA employs the J (1.24$\mu$m), H (1.66$\mu$m) 
and  K$_s$ (2.16$\mu$m) bands. The VVV CA  combines near IR bands and  we connected it to  Aladin 9.0.  The WISE CA  includes  the   
W4 (22 $\mu$m), W2  (4.6$\mu$m) and  W1 (3.4 $\mu$m) bands, and is sensitive to stars and dust emission. The Spitzer CA consists  
of  a set  of bands from 3.6$\mu$m (IRAC1) to 8.0$\mu$m (IRAC4).  Herschel uses the  70$\mu$m and 100$\mu$m bands  for  a BR color 
composition representing  ESA's  Photodetector Array Camera and Spectrometer (PACS)  CA, which is sensitive to relatively cold dust 
emission and  dust caps around YSOs. The DSS, 2MASS and  WISE surveys generated all-sky atlases. VVV is dedicated to the bulge/central 
disk region, while   Spitzer  and Herschel have particular coverages, especially along the disk.

Fig.~\ref{f1} shows the new OC candidate (OCC) Bc\,105.  In the optical with the DSS  RGB it is not detected owing to contamination 
and absorption effects. It shows up in the near IR with 2MASS. In the mid IR (WISE) it shows no dust emission, as expected for an OC. 
Possibly, many of the new OCCs will require analyses with deeper photometry than 2MASS.

Figure~\ref{f2} shows the EC Bc\,74 (Ryu\,265 - see Sect.~\ref{sec:5}) which is invisible in  DSS  owing to high absorption. In 2MASS it becomes 
a conspicuous cluster. Many clusters from \citet{Froebrichetal2007} present this behavior.  Dust emission shows 
up in WISE, revealing an associated EC. It also shows cold dust emission in the far IR (Herschel).

Figure~\ref{f3} shows the EC  Bc\,37 (Ryu\,675). It is undetected in DSS and very contaminated in 2MASS.  Spitzer  reveals 
a prominent deeply embedded EC  with diffuse dust emission and dust-capped YSOs. Finally, Herschel  shows strong 
cold dust emission.


Table~\ref{tab4} deals with kinematical groups and kinematical associations.  The latter are nearby young stellar 
groups  that may cover large solid angles in the sky, or even have an all-sky distribution, if the Sun is located 
within their volumes (e.g. the Local Association - \citet{Zhao2009}. Table~\ref{tab5} compiles  Local Group galaxies 
and indicates their star cluster and association content.

The classification scheme in itself is for the first time presented in the literature concerning a general catalog. 
The first approach is to adopt the original author's classification, but in dubious cases, we re-analyzed the object 
by means of morphological properties (Sect.~\ref{sec:2.1}) in images and the available information in the literature. When a new result 
supersedes the original classification(s), we adopted it, e.g. the case of 5 cluster remnant candidates that 4 of them 
turned out to be asterisms with Gaia (\citealt{Kos2018}).

\subsection{The Subcatalog of Newly Found 638 Entries}
\label{sec:3.2}

During the last 20 years many new SCAOs  were identified in the surveys by one of us (E.B.). A number of them have 
in the mean time been published by other authors, and were excluded. We ended up with 661 new objects. They were  
subsequently merged into  CatClu.  ECs are frequent  in this list owing to  evident dust emission in  WISE  
(\citealt{CBB2015a}, \citealt{CBB2016a}), as well in Spitzer and Herschel, when available. The discoveries
amount to 638, because 23 are in common with \citet{wise1} - see Sect.~\ref{sec:5}.

\subsection{The CatClu catalog}
\label{sec:3.3}

We retrieved the chronological order of publications in the literature. Re-discoveries are not a demerit because new 
detections of a given object will support its existence, in particular if it shows up   in different wavelength ranges. 
Electronic Table~\ref{tab3} shows star clusters, candidates and alike objects. By columns: (1) and (2) Galactic, (3) and 
(4) J2000.0 equatorial coordinates, (5) and (6) major and minor diameters in arcmin, (7) classification code, 
(8) chronologically ordered cross-identified designations, (9) long  field for comments, and (10) reference code. 
 

\begin{deluxetable*}{lrrrrrllcl}
\tablecaption{Clusters, Candidates and Alike Objects\label{tab3}}
\tablehead{
\colhead{l$_G$}&\colhead{b$_G$}&\colhead{$\alpha$}&\colhead{$\delta$}&\colhead{D}&\colhead{d}&\colhead{Class}&\colhead{Name/Acronym}&\colhead{Comments}&\colhead{Ref.} \\
\colhead{($^o$)}&\colhead{($^o$)}&\colhead{(h:m:s)}&\colhead{$(^o:\arcmin:\arcsec)$}&\colhead{($\arcmin$)}&\colhead{($\arcmin$)}&\colhead{}&\colhead{}&\colhead{}&\colhead{Code}
}
\colnumbers
\startdata
0.00  & -1.30   & 17:50:43  & -29:36:22  & 4    & 4    & GCC  & Minni\,40,Minniti\,40 & based on RR Lyrae, A$_{K}$=0.46  & 3404\\
0.01  & -1.90   & 17:53:09  & -29:54:20  & 2    & 1.5  & OC   & AL\,2,Andrews-Lindsay 2,        & $-$ & 213 \\
&&&&&&&ESO\,456-5,MWSC\,2734&&\\
0.02  & -1.43   & 17:51:17  & -29:39:04  & 1.0  & 1.0  & OC   & VVV\,51,VVV\,CL151  & prominent  & 1000 \\
0.02  & -1.34   & 17:50:56  & -29:36:39  & 4    & 4    & GCC  & Minni\,56,Minniti\,56 & based on  RR Lyrae $ A_K$=0.46   &3404   \\                                                
0.03  & -0.29   & 17:46:51  & -29:03:47  & 1.5  & 0.9  & OCC  & DB\,1,Dutra-Bica\,1        & $-$ & 098,126\\
\enddata
\end{deluxetable*}

Some of these class samples have increased dramatically, in particular ECs, lPOCRs and asterisms. Cataloging SCAOs dates 
back to at least  2  centuries with Messier's list. CatClu  shows  small and/or fainter clusters, in general more absorbed 
than larger more populated classical ones in  WEBDA, DAMLO2  and MWSC.  The former two deal  with  OCs with parameters collected 
from  the literature. MWSC derived parameters for 3006 OCs, GCs and associations. However,
they analyzed additionally 778 objects that were considered to be (i) not a cluster, (ii) possible cluster but
parameters not determined, and (iii) duplications. We checked all these classifications and incorporated them
to the catalog. We agree that many are not clusters or alike, but a considerable fraction is of embedded clusters
owing to the dust emission now seen in WISE, Spitzer or Herschel.

Vizier provides catalogs but does not cross-identifies them, and does not deal with small samples or individual clusters in 
papers.  SIMBAD shows cross-identifications but in general  does not consider chronology. The present work overcomes these 
limitations.  This  multi-band survey (Sect.~\ref{sec:3.1}) is  a powerful tool to revise, detect  and classify objects in 
the optical and IR.  We also find and/or compile poorly populated objects like compact  and loose POCRs \citep{PavaniBica2007}. 
They are a fundamental sample to be explored, probably dealing with aspects of cluster dynamical evolution and dissolution. 
The present catalog is both  a base of objects for {\em Gaia} and a  database for further developments in SIMBAD and Vizier. 
Some ECs seen in WISE and Spitzer are so much  absorbed that they are not expected to be in {\em Gaia}. A discussion  on  
angular distributions of different object classes and their  total populations is provided in Sect.~\ref{sec:4}. Table~\ref{tab1} 
shows  that to date ECs outnumber  OCs.  The analyses of asterisms may  reveal  a number of star clusters. In particular, we 
detected several ECs among the asterisms. About 7700 entries in CatClu require first studies. 

We emphasize that such entries are not only cluster candidates. According to
Tab.~\ref{tab1} we have important populations of e.g. OCCs, ECCs and GCCs. Thus, our classification
is a step further in terms of class discrimination. It is important to remark that for some authors a
cluster only becomes so after astrophysical parameter determination. As a consequence, there are many
obvious clusters (identified morphologically) that are not studied in terms of astrophysical
parameters; e.g., among the 4234 ECs, only a few hundred have parameters. 
 
\subsection{Our Group Contributions}
\label{sec:3.4}

Our group contributions to the general catalog started as lists developed with  2MASS, e.g. \citet{DutraBica2000} with 58, 
\citet{Dutraetal2003} with 179, and  \citet{Bicaetal2003a} with 167 entries. More recently, \citet{CBB2016a} and references 
therein provided 1101 entries, mostly ECs found in WISE. Several of our studies included a few clusters. In all, we published 
28 papers contributing with  2336 entries ($21\%$) of Table~\ref{tab3}. Taking into account  cross-identifications and 
chronological ordering of designations, our group has discovered $\sim20\%$ of CatClu. In addition, two of us (C.B. and E.B.) 
collaborated in  cluster lists of  the VVV Survey (\citealt{Borissovaetal2011} and \citealt{Borissovaetal2014}). Finally, 
\citet{Bicaetal2003b} compiled  ECs  and embedded groups from the IR literature,  which are now incorporated in 
Table~\ref{tab3}.

\section{Angular Distributions and Statistical Properties}
\label{sec:4}

\begin{figure}
\plotone{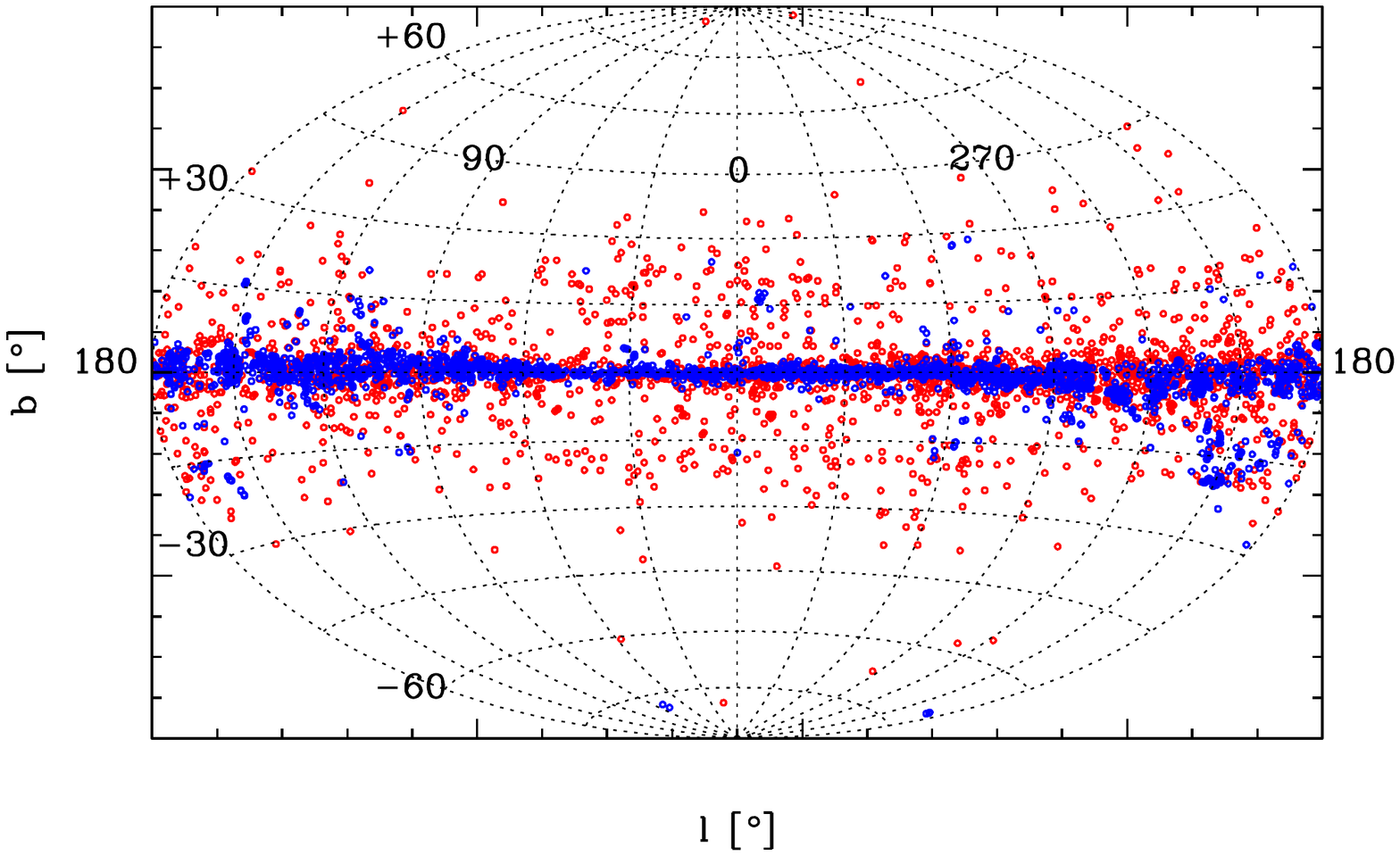}
\caption{Aitoff projection for the ECs (blue circles) compared to OCs (red circles).}
\label{f4}
\end{figure}

\begin{figure}
\plotone{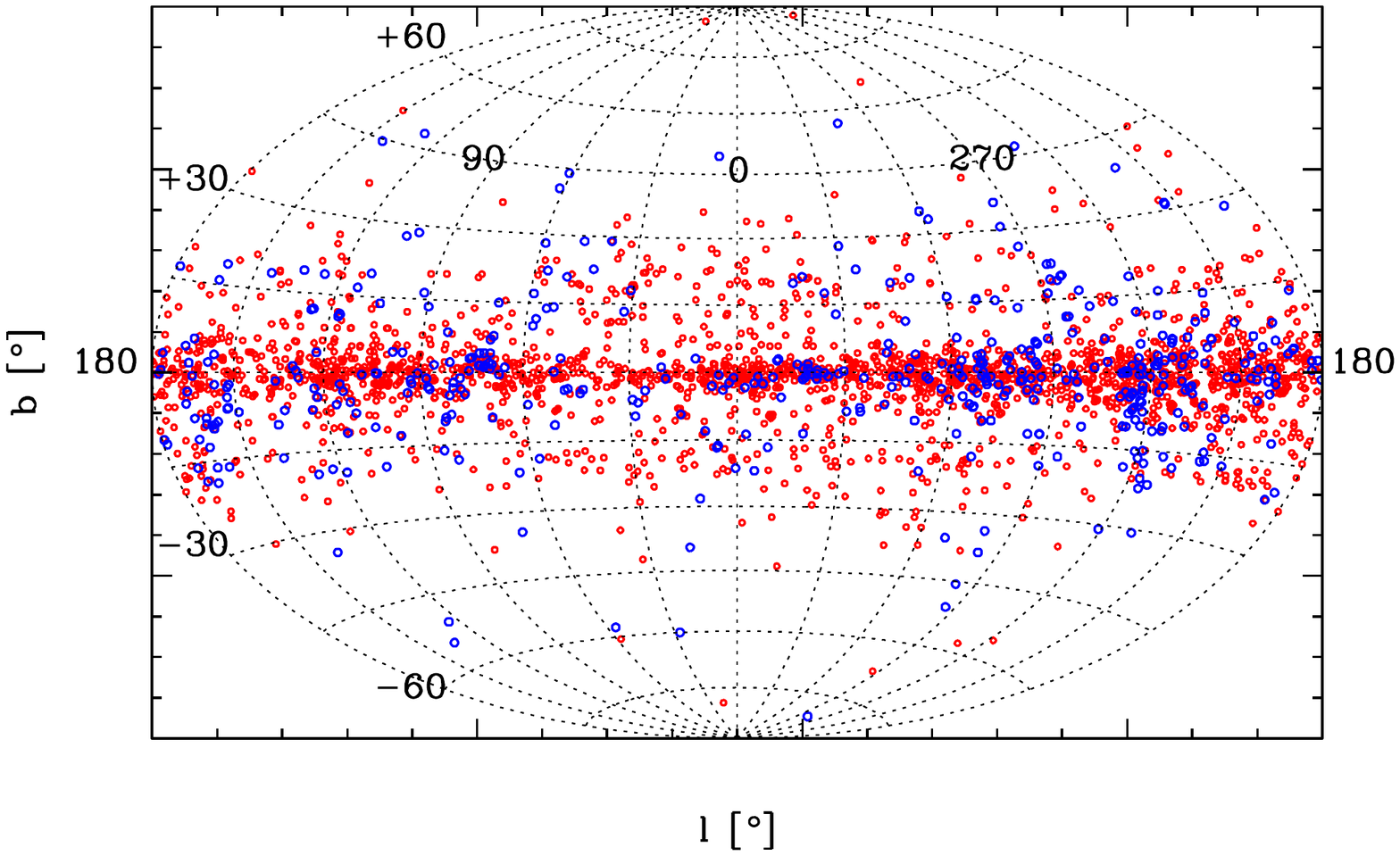}
\caption{Same as Fig.~\ref{f4}, but for lPOCRs (blue circles) compared to OCs (red circles). }
\label{f5}
\end{figure}

\begin{figure}
\plotone{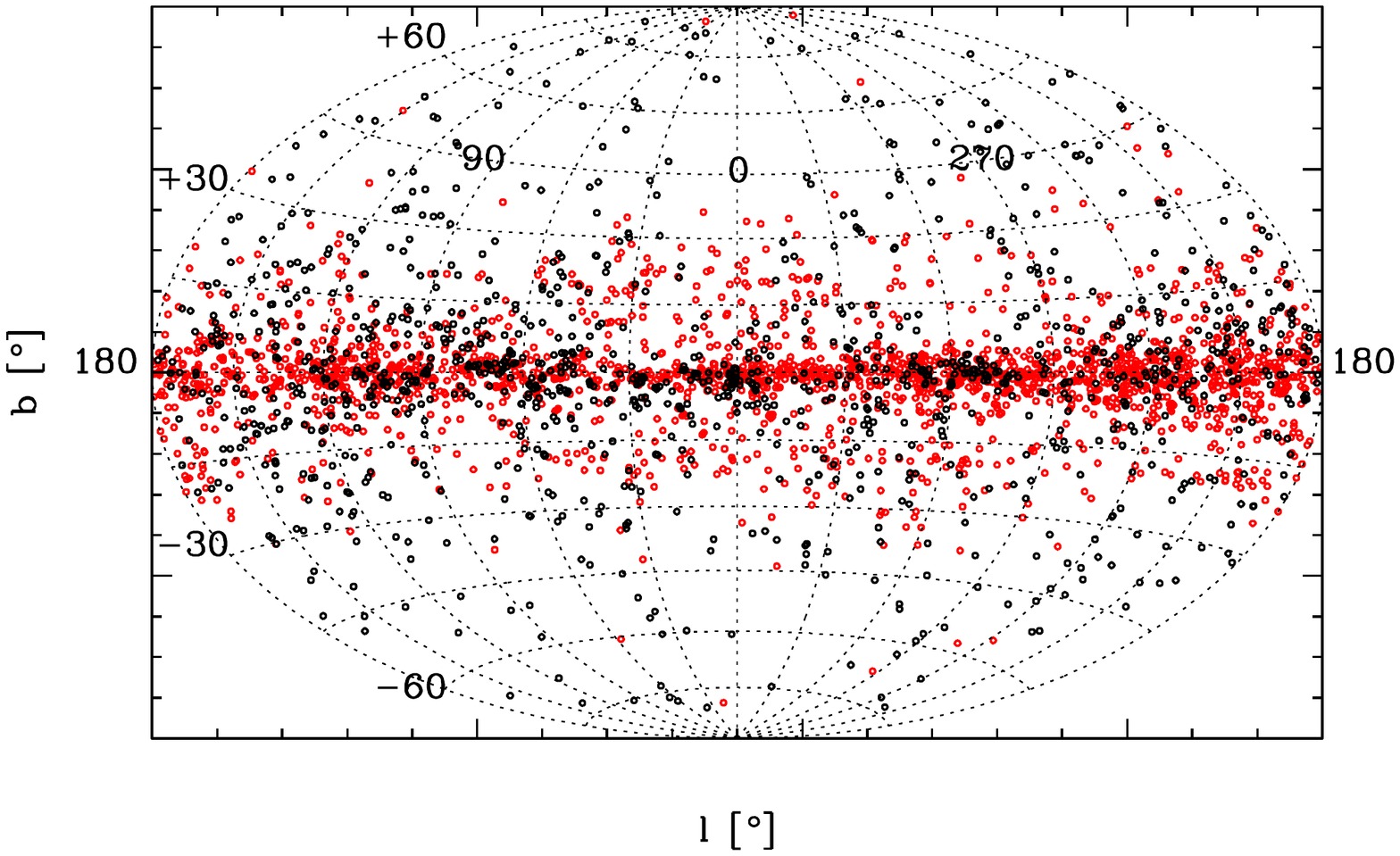}
\caption{Same as Fig.~\ref{f4}, but for Asterisms (black circles) compared to OCs (red circles). }
\label{f6}
\end{figure}

\begin{figure}
\plotone{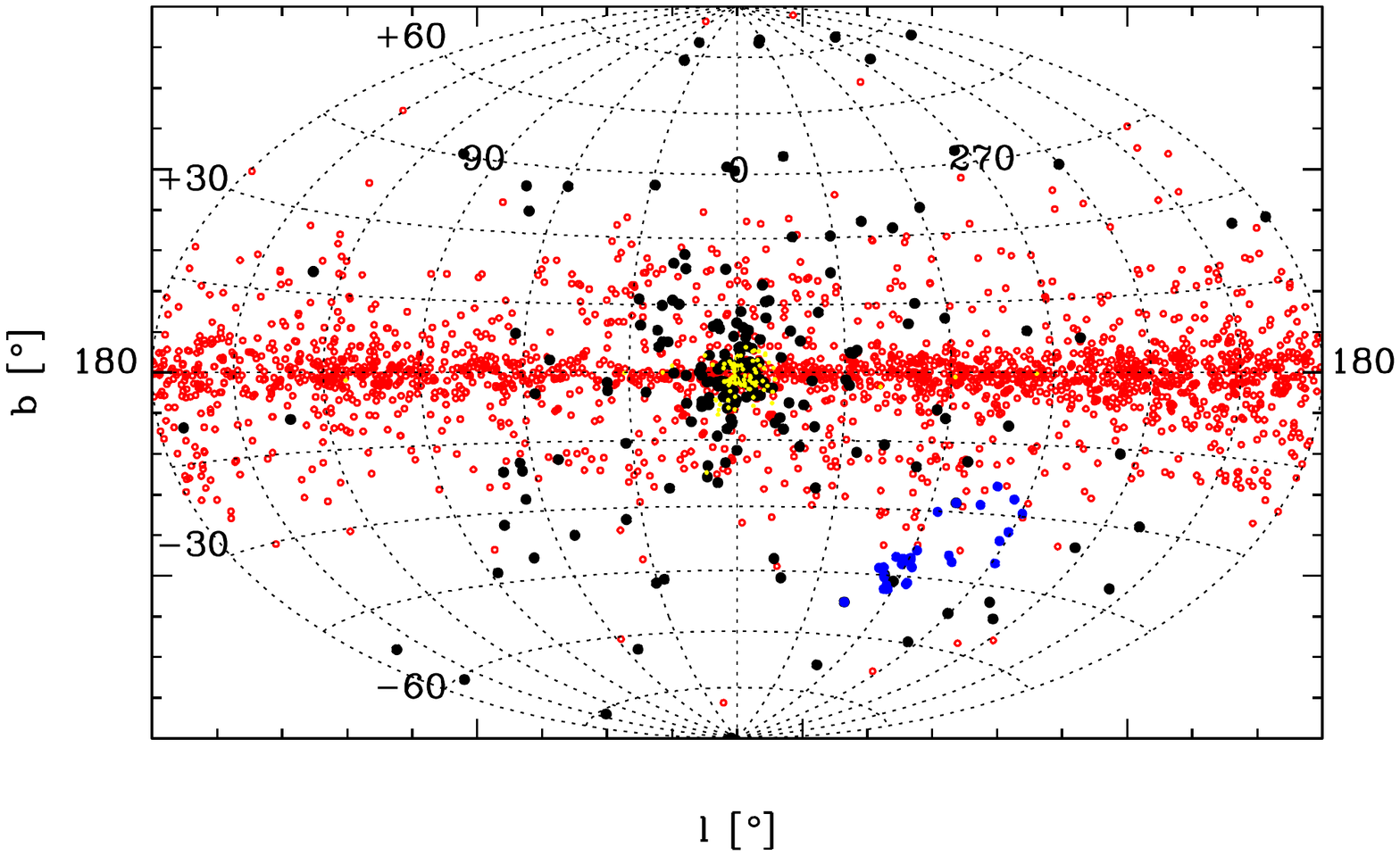}
\caption{Same as Fig. 4, but for GCs (black dots) and GCCs (yellow dots), compared to OCs (red circles). 
Also shown are MHC clusters (blue circles).}
\label{f7}
\end{figure}

\begin{figure}
\plotone{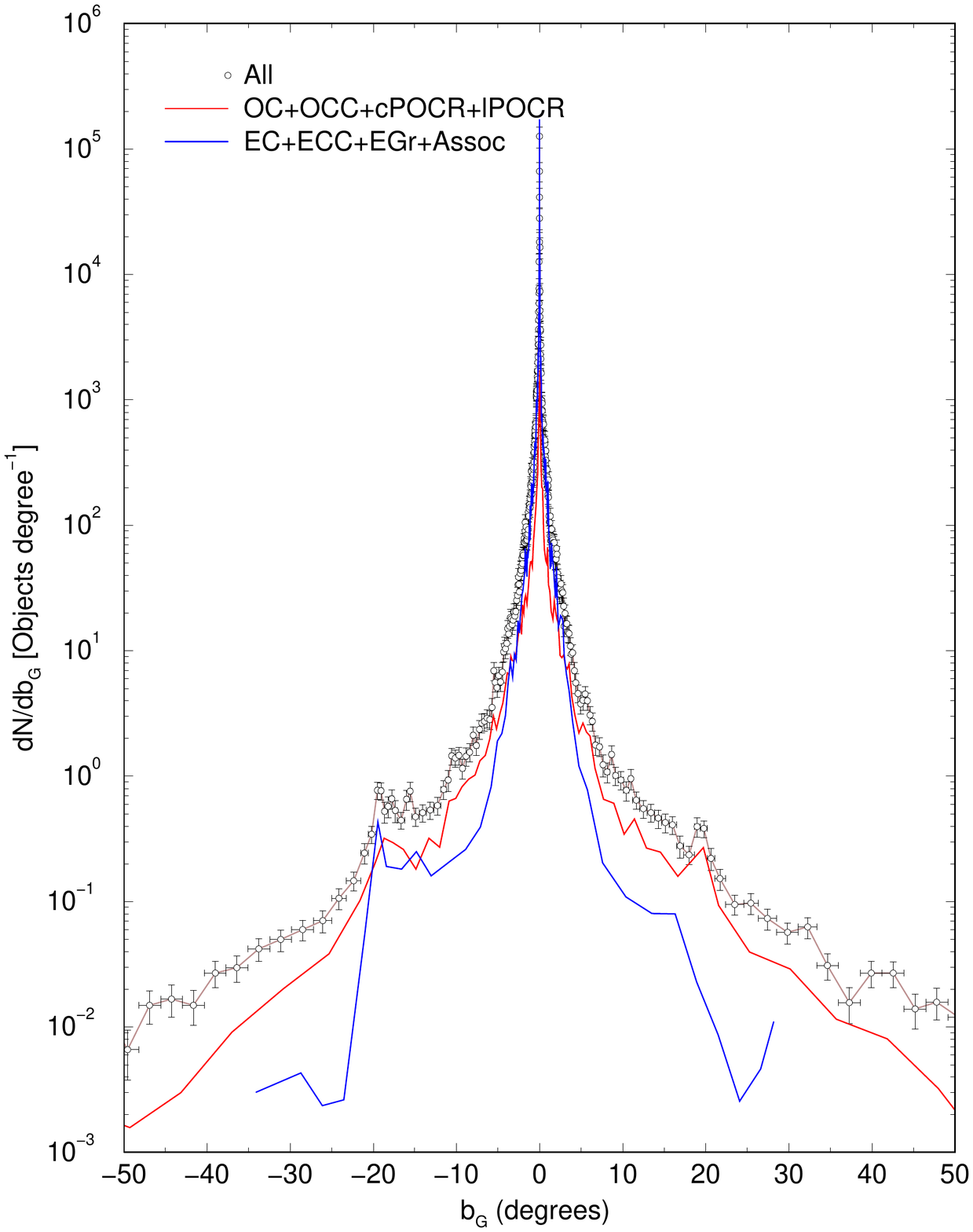}
\caption{Galactic latitude ($b_G$) distribution functions normalized for peak value. Star-forming samples 
(EC+ECC+EGr+Assoc) compared to evolved ones (OC+OCC+cPOCR+lPOCR), and with the overall sample (excluding 
asterism and MHCs. }
\label{f8}
\end{figure}

\begin{figure}
\plotone{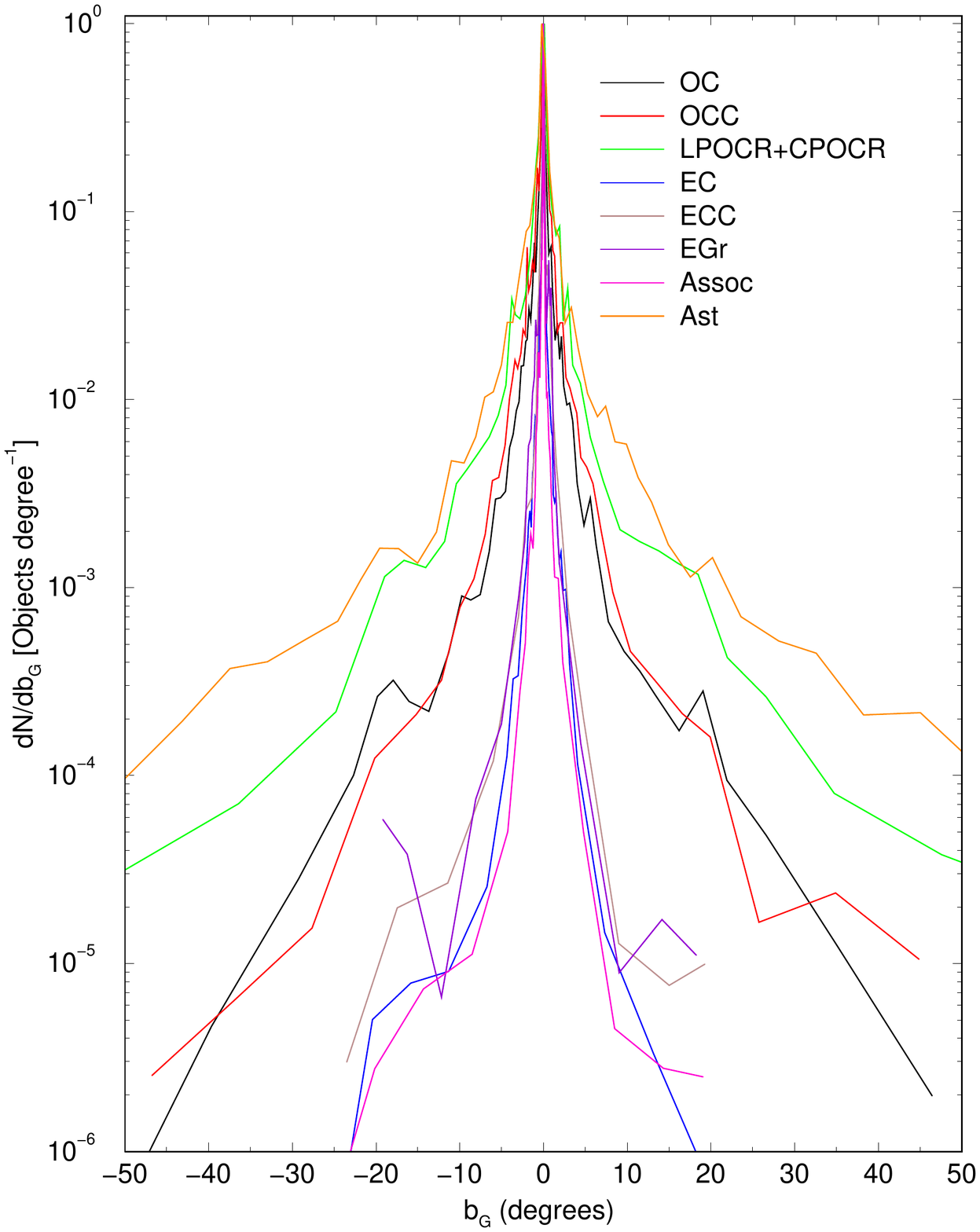}
\caption{Same as Fig.~\ref{f8}, but for individual samples.}
\label{f9}
\end{figure}

\begin{figure}
\plotone{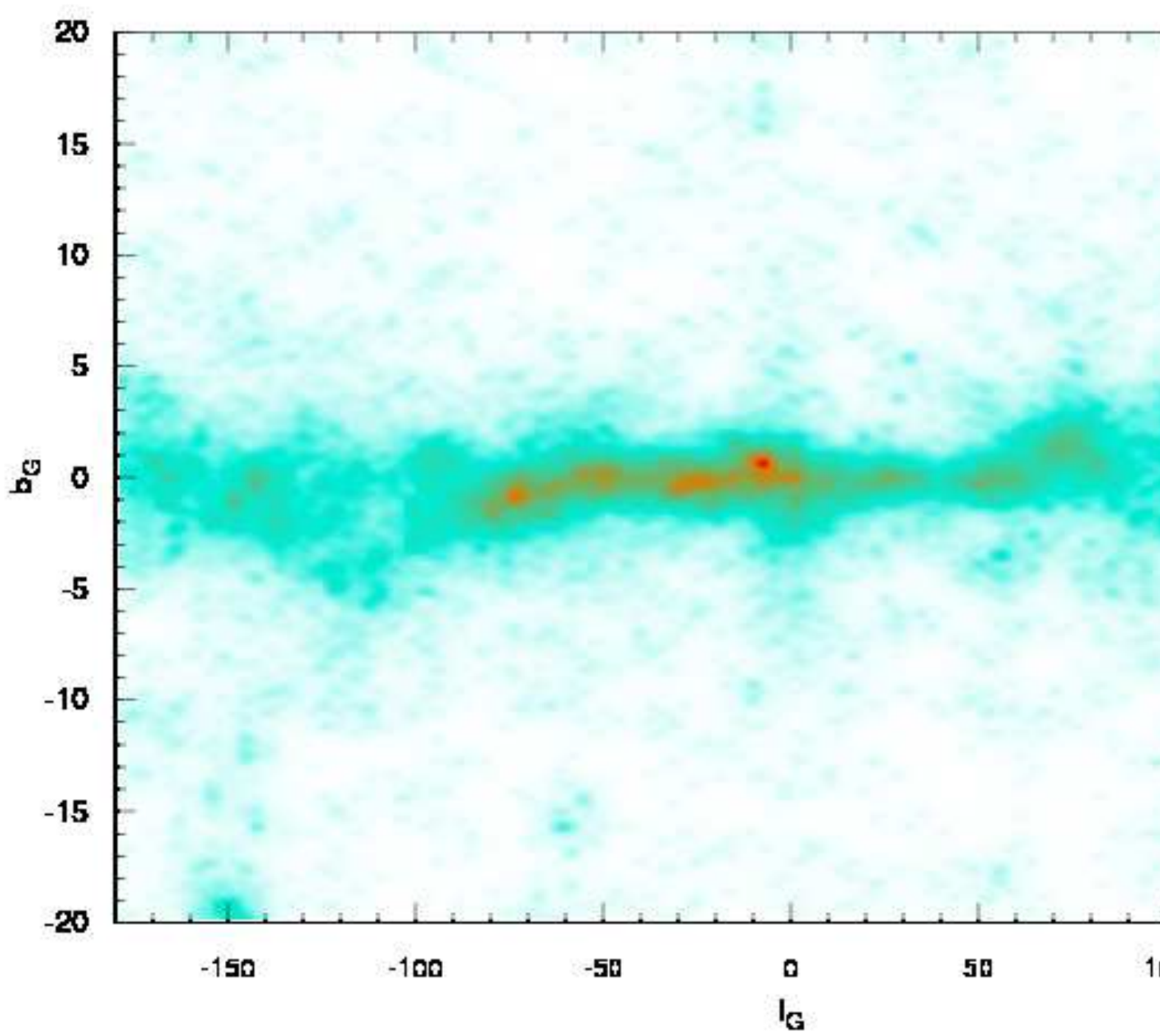}
\caption{Number-density map (in Galactic coordinates) of the total sample (excluding the Asterisms).}
\label{f10}
\end{figure}

\begin{figure}
\plotone{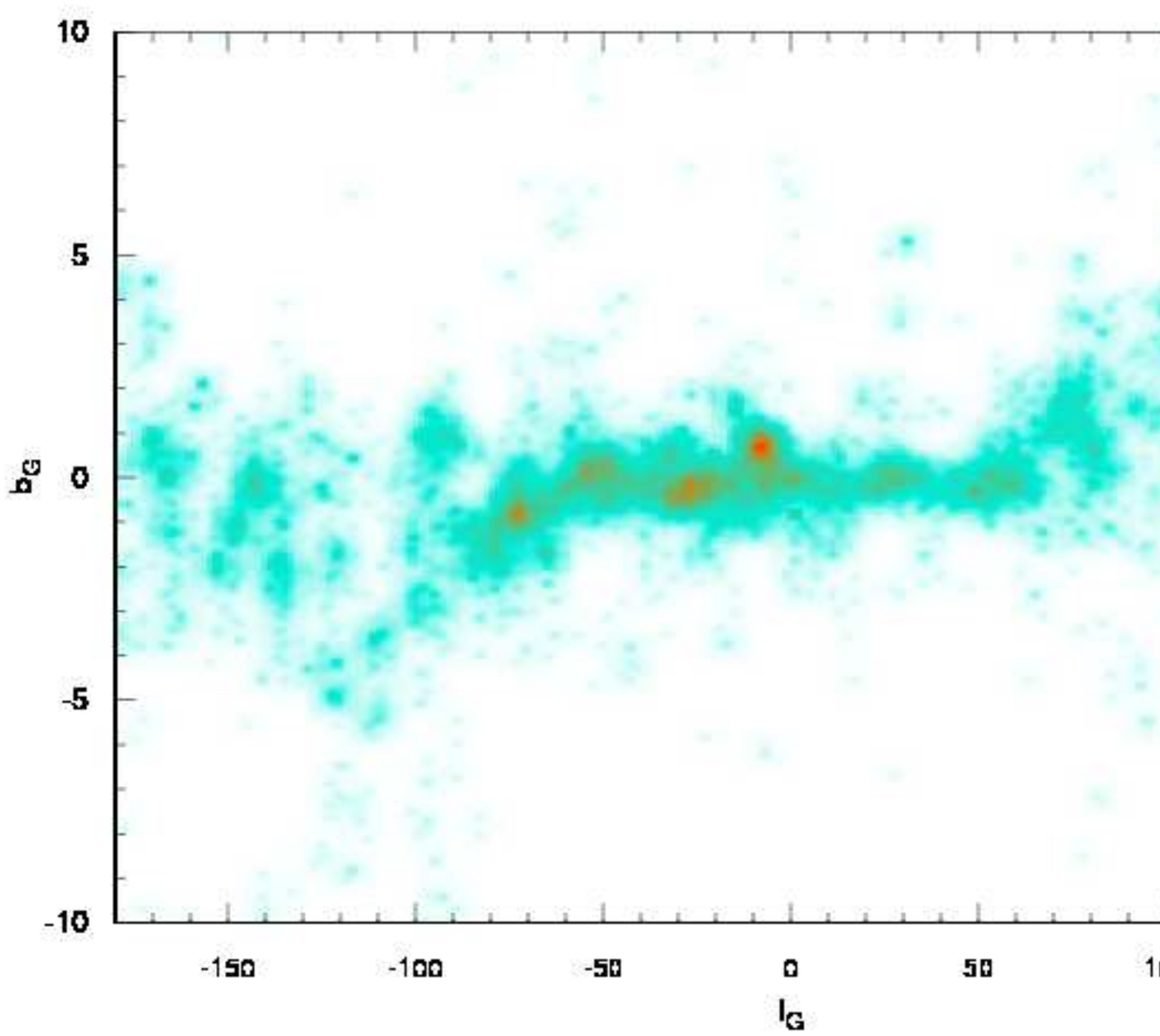}
\caption{Same as Fig.~\ref{f10} for the EC sample, thus enhancing the warp and flare. }
\label{f11}
\end{figure}

\begin{figure}
\plotone{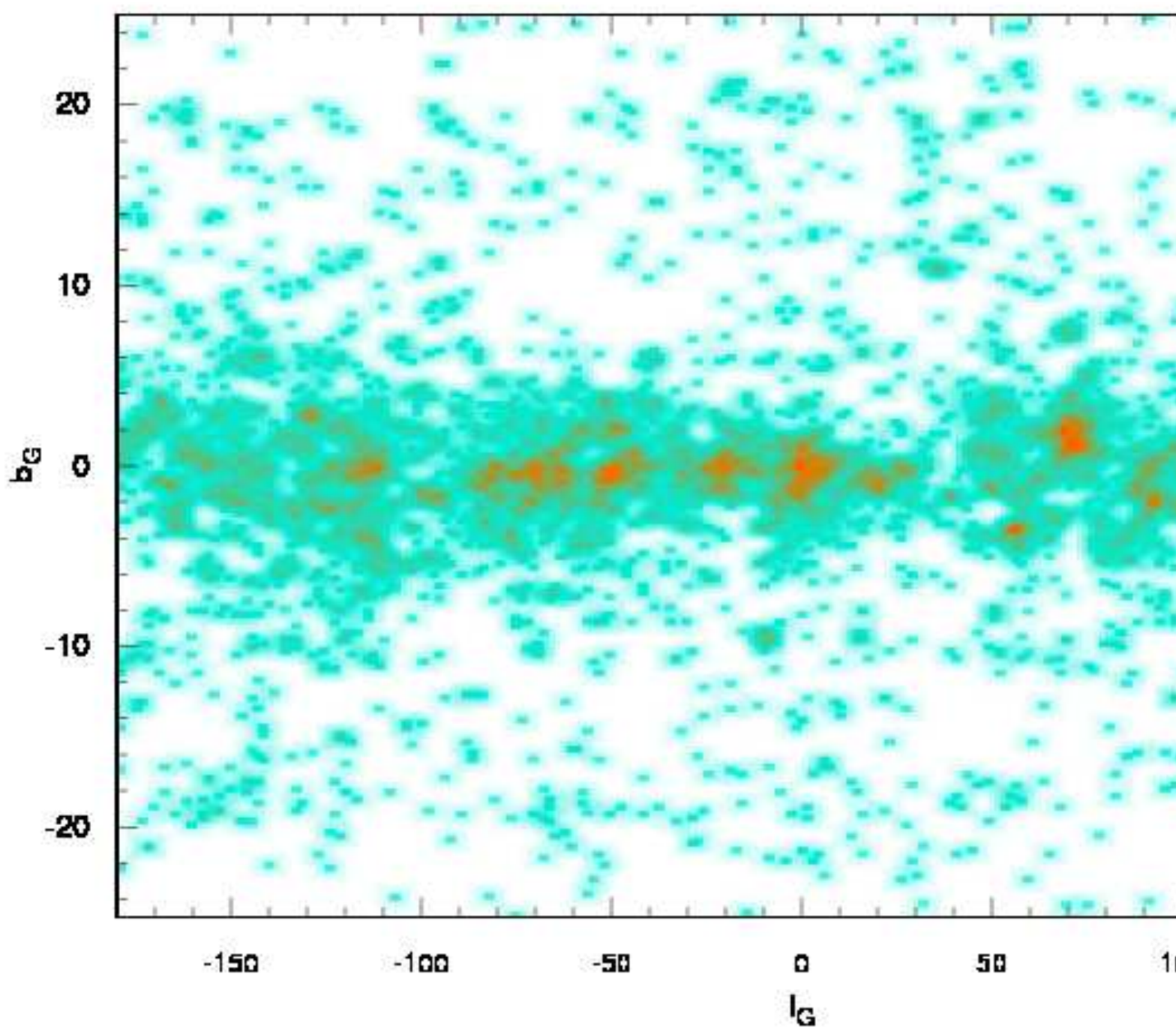}
\caption{Same as Fig.~\ref{f10} for the OC sample, showing overdensities. }
\label{f12}
\end{figure}

Statistical properties of the object sample and subsamples are considered using angular distributions  
of the object types. We employ: (i) Galactic coordinates $l_G \times b_G$ in Aitoff projections for a 
selection of  object types in CatClu; (ii) $b_G$ distribution functions to compare the overall structures; 
(iii) number-density maps to check for internal structures. 
Figures~\ref{f4} to \ref{f12} do not include the new discoveries in the literature, because
they have an impact essentially in the central parts (Sect.~\ref{sec:5}).

Figure~\ref{f4} shows the large EC sample  compared to OCs. The bulk of 
the ECs is more tightly distributed to the plane. Some ECs are projected at intermediate latitudes and a few 
appear to be related to gas and dust  halo clouds (\citealt{Camargoetal2015b}, \citealt{CBB2016b}). The bulk 
of OCs attains higher latitudes than that of ECs and may suggest that OCs often become dynamically heated 
with time. Alternatively, they might acquire their orbital properties preferentially at birth in the plane.

Figure~\ref{f5} shows the lPOCR sample superimposed on OCs. The distributions are 
similar, although the lPOCRs attain on the average somewhat higher latitudes (Fig.~\ref{f9}). However, the 
overlapping region suggests that  OCs and POCRs appear to be statistically related, supposedly in terms of 
evolutionary terms.

Asterisms (Fig.~\ref{f6}) are more widely distributed in $b_G$ than OCs. The fraction of the distribution 
in common suggests the presence of  physical objects in the asterism sample, while the high latitude excess 
might imply  contamination  of chance stellar concentrations.

The GC population has recently had an important progress (Fig.~\ref{f7}). \citet{Bicaetal2016} have 
decontaminated the bulge of halo intruders. The confirmed  population is about to reach 200 GCs 
\citep{Minniti2017b}. GCCs in the bulge area have also increased, especially in the VVV area 
\citep{Minniti2017a}. The bulge GCs are strongly concentrated to the center, while the halo ones populate 
most of the celestial sphere at low densities.

Several GCs trace the  Milky Way outer halo beyond 100 kpc \citep{Harris2010}. The LMC and SMC cluster systems, 
with heliocentric distances of 51 and 64 kpc \citep{McConnachie2012}, respectively, are currently enclosed in 
the Galactic halo. SMC and LMC  halo clusters are detached from their respective main bodies \citep{Bicaetal2008}. 
Recently, the clusters SMASH\,1 \citep{Martinetal2016}, {\em Gaia} 3, Torrealba\,1, DES\,4 and  DES\,5 
\citep{Torrealbaetal2018} were detected in the LMC halo, while Tuc V is probably a cluster or remnant far in the 
SMC halo. All outlying LMC clusters in \citet{Siteketal2016} are projected on the  LMC disk, while 8 SMC outliers 
in \citet{Siteketal2017} can be classified as SMC halo's. Such clusters (Fig.~\ref{f7}) may be eventually 
accreted by the Milky Way, and thus we introduce the Magellanic Halos's Clusters (MHC) class in Table~\ref{tab1} 
and Table~\ref{tab3}. The interaction between the LMC and SMC led the SMC to tidal stripping \citep{Diasetal2016}, 
favouring the occurrence of SMC halo clusters. The SMC now has 22 and the LMC 11 cataloged clusters in their halos, 
and the angular distribution is given in Fig.~\ref{f7}. We point out that in the present census we exclude young 
clusters related to the tidal Bridge that connects the Clouds (\citealt{Bicaetal2008}, \citealt{Bicaetal2015}). 
At any rate, they populate the Galactic halo with a young stellar component.


The total star-forming sample (Fig.~\ref{f8}) has a narrower distribution than the evolved one. The former 
distribution also decays fast for $|b_G| > 35^o$. This clearly shows how the projected star-formation is 
concentrated to the plane. A small local excess occurs especially in the evolved sample at $|b_G|\approx20^o$.  
Such features may correspond to changes in the OC population \citep{Gabriel2017}.

It is remarkable how all star-forming subsamples present essentially the same $|b_G|$ distribution (Fig.~\ref{f9}), which 
tend to form near to the plane. OCs and the large sample of OCCs have comparable widths, suggesting that the OCCs are worth 
data-mining  for OCs.   Finally, POCRs depart  somewhat from OCs, and asterisms depart yet more. However, both POCRs and 
asterisms overlap major fractions of their distributions with OCs, again indicating the need of data-mining.


Fig.~\ref{f10} shows the Galactic warp for $|l_G|<100^o$, together with the disk flares for $|l_G|>100^o$ (\citealt{Momanyetal2006} 
and references therein). The warp is clearer in  the  EC sample (Fig.~ \ref{f11}), together with a flare effect.

In the OC distribution (Fig.~\ref{f12}) a series of small overdensities occurs along the plane. They have some counterparts in 
the Galactic spiral arms and their tangent points \citep{Vall2017}. Tangent points imply line-of-sight accumulation of HI, HII, 
dust and young clusters. The occurrence of excesses in tangent points might be explained by non-embedded young clusters in the 
range 10-50 or even as old as 100 Myr. We caution for the possibility of over or undersampled zones, since the results stem 
from many authors. Nevertheless, the present cross-identifications minimized such effects.

\begin{figure}
\plotone{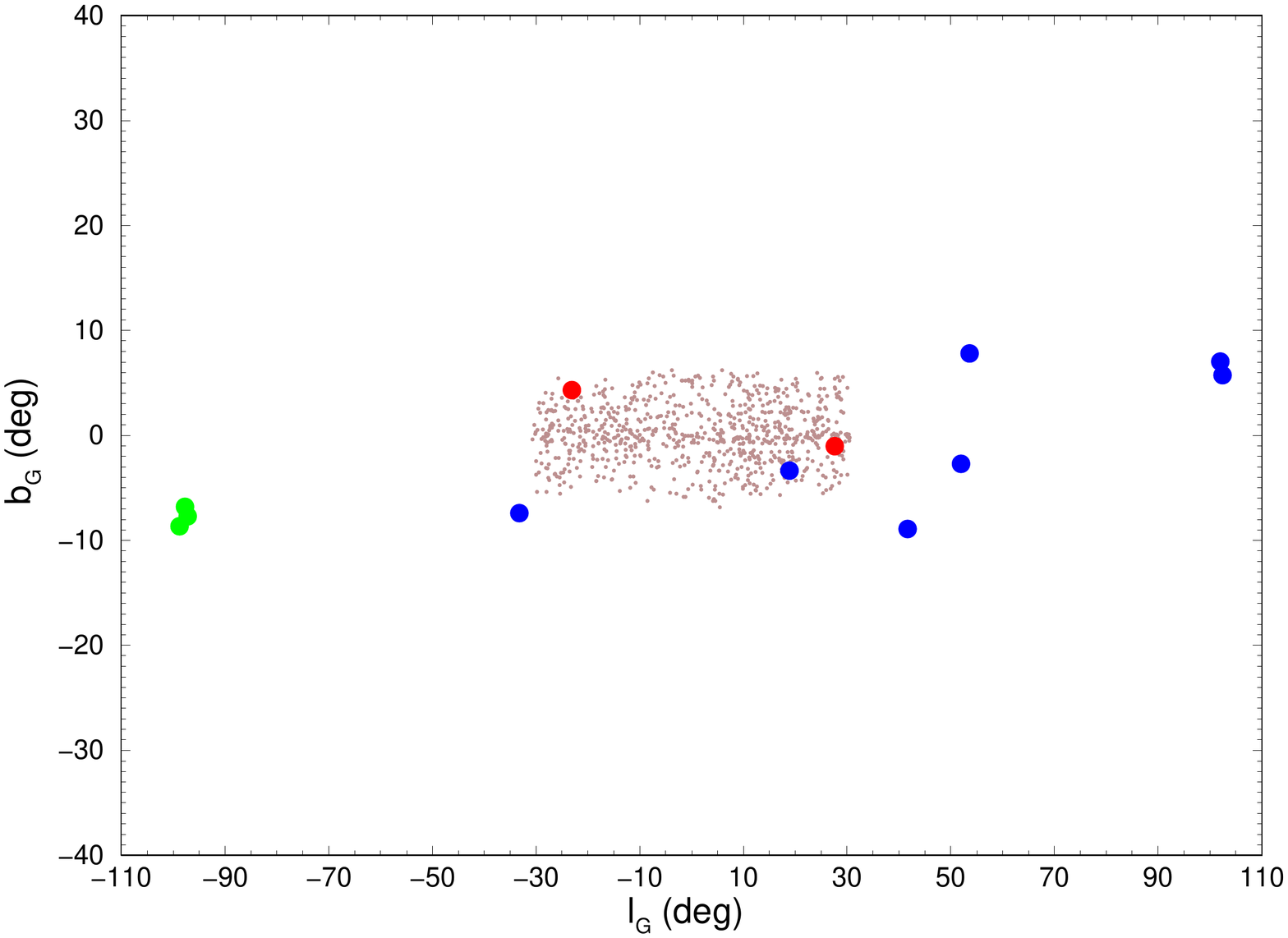}
\caption{The recent additions from literature: \citet{wise1} (brown symbols); the 2 new GCs of \citet{wise2} 
(red circles); \citet{gaia2} (green circles); \citet{gaia1} (blue circles).}
\label{f13}
\end{figure}

\section{Concluding Remarks}
\label{sec:5}

The present work was carried out essentially in an independent way of SIMBAD. This was achieved both for 
cross-identifications of large lists, such as FSR \citep{Froebrichetal2007} and MWSC (\citealt{Kharchenkoetal2013}), and   
individual object  papers. It would be important  to keep the present SCAO catalog undismembered in  VIZIER and/or CDS.    

It would also  be important  that SIMBAD incorporates  the present results taking into account  the unique efforts   employed 
here to identify and characterize SCAOs, and shed light in cases of  doubt.  

We further stress the importance of the different object classes, from ECs, through OCs to remnant candidates (\citealt{Pavanietal2011}; 
\citealt{Bicaetal2001}). Different evolutionary paths can connect  them both photometrically and structurally. In particular, ECs can 
dissolve,  never evolving to open clusters \citep{Lada2003}, and  they might produce as well young cluster remnants.  On the other hand,  
asterisms have proven to be very  useful for  data-mining \citep{BicaBonatto2011}. 

The number of GCs and candidates in the MW has increased dramatically to 294 (Tables~\ref{tab1} and \ref{tab3}) 
as compared to the 157 GCs listed by \citet{Harris2010}. This progress was achieved mostly by studies unveiling 
the low-luminosity bulge GCs, e.g. \citet{Minniti2017a}, \citet{Piatti2018}, \citet{BicaMinBon2018}, and 
\citet{Camargo2018}. Considering  the MW GC candidates, this can potentially alleviate the MW count deficiency 
with respect to M\,31, with 361 GCs in the sample of \citet{CaldwellRomanowsky2016} and the total estimate of 
$\approx450$  by \citet{Larsen2016}, recalling that the MW and M\,31 appear to present  comparable masses 
(\citealt{PhelpsNusserDes2013}).

SIMBAD has been successful in  cross-identifications of point sources like stars among  many  catalogs, since they are more suitable  
for  automated tools. Diffuse distant galaxies close to the  point source limit are likewise well suited for   cross-identifications 
among large databases  e.g.  SIMBAD and the NASA Extragalactic Database. On the other hand,  SCAOs are extended, sometimes angularly  
large, with low surface brightness variations owing to stellar depletions, contamination, absorption and  distance effects, as well 
as  dependent on  observational and instrumental conditions  in a given  survey. They may  show neighbors or hierarchical distribution.  
Care is necessary in handling them. Automated searches of overdensities have provided new clusters.  Automated analyses of CMDs and 
cluster structure  have given, in turn, fundamental parameters (\citealt{Kharchenkoetal2013}).  Samples detecting new clusters by searching   
overdensities and deriving their parameters were recently obtained (\citealt{Schmejaetal2014},  \citealt{Oliveiraetal2018}). 

Awaiting for automated future searches  and analyses in any survey environment, the present work has carried out the  task of  
homogeneously  organizing   all previously available and  new  SCAOs into a single database. It furnishes  a collection of several 
thousand fresh objects for photometric and  spectroscopic  studies, as  {\em Gaia} unveils   parallaxes and   PMs (\citealt{Gaia2017}).  
We have developed   a database of  3 catalogs:  (i) clusters, candidates and alike objects (Table 3), (ii)  kinematic groups and 
kinematic associations (Table~\ref{tab4}), and (iii)  Local Group galaxies and their star clusters and associations (Table~\ref{tab5}). 
These  catalogs   share  the same list of  792 references. Table~\ref{tab3} with 10978 star clusters and  candidates outnumber by 
several thousands the entries in any previous catalog or database. 

{\em Gaia} will give precise parallaxes, PMs  and photometric measurements for several hundred  nearby open clusters, providing 
essential constraints to stellar evolution models (\citealt{Randichetal2018}, \citealt{Gaia2017}).  For star  clusters and stellar 
groups with increasing distance from the Sun the 3D distribution of different cluster generations will be obtained, and within 
uncertainties, probably that of the entire near side of the Galaxy. Undoubtedly, the present effort, with thousands of entries 
more than any previous catalog or database,  is  a timely contribution to  {\em Gaia}.

As this work approached completion, 2 papers with WISE (\citealt{wise1}; \citealt{wise2}) and 2 others with 
{\em Gaia} DR2 (\citealt{gaia1}; \citealt{gaia2}) data have been published with new discoveries that add to the 
present work. Concerning both WISE results, the authors have discovered hundreds of new clusters, most of them 
non-embedded, including 2 new Globular clusters. The {\em Gaia} works represent the first wave of discoveries with
{\em Gaia} DR2 (\citealt{Gaia2018}). In this context, the present work is a valuable tool to minimize re-discoveries
and facilitate the search and identification of targets. Finally, we remark that the new discoveries above have been
fully cross-checked and the new ones incorporated into the catalog. The already existing objects were indicated as
equivalent in the list of designations in the respective catalog.

Given the accuracy of the coordinates in the present catalog, as well as those in \citet{wise1} for 921 clusters
(202 ECs and 719 OCs), 
we applied for the first time in this study a  position matching routine. For separations larger than 120\arcsec, the 
catalogs have no objects in common for 848 clusters, which are thus new. The catalogs match for 73 clusters: (i) we 
confirm 24 as additional new clusters as a rule in pairs,  while (ii) 26 coincide  with the previous literature, and 
finally, (iii) 23 are equal to clusters found by one of us (E.B. - the present Bc clusters), including Ryu\,265 = Bc\,74 
(Fig.~\ref{f2}), Ryu\,675 = Bc\,37 (Fig.~\ref{f3}). Thus, in Table~\ref{tab3} we assigned them Ryu as the first 
designation, since the publication date establishes a discovery (Sect.~2.4). The Bc designations were mantained in 
Table~\ref{tab3}, as well as in Figs.~\ref{f2} and \ref{f3}, as additional ones, in the sense that detections by 
different authors give more weight to a given cluster. The new additions are shown in Fig.~\ref{f13}. The Gaia
discoveries so far deal with nearby clusters along the disk, while the WISE sample roughly corresponds to a
rectangular distribution in the central parts. We checked the impact of the new objects on Figs.~\ref{f11} and 
\ref{f12}, finding that the central sample produces a slight enhancement of the structures in both figures. The 
rectangular shape (Fig.~\ref{f13}) persists for OCs, but essentially merges into the EC sample.

A website containing the present database will be made available soon. We emphasize that the online database will
be updated as new entries appear.

\acknowledgements
We thank the anonymous referee for important comments and suggestions. We thank Dr. A. Pieres
for pointing out recent literature on Galactic streams.
We acknowledge the use of Aladin and IPAC tools, together with the surveys therein. C.B. and  
E.B. thank support from the Brazilian Council CNPq.

\appendix

\section{GALACTIC KINEMATICAL GROUPS AND CLUSTERS IN LOCAL GROUP GALAXIES}

Two additional catalogs were compiled in this work: (i)    245 kinematical groups  (CatKGr, Table~\ref{tab4}),  and (ii)  
144 Local Group galaxies and their cluster and/or association content (CatGal, Table~\ref{tab5}).  CatKGr  shows objects  
that do not fit OC characterizations, such as  moving groups, and the fast  growing   domain of  halo streams. CatGal    
provides references to  star clusters and associations in Local Group (LG) galaxies. It also includes the  ultra faint 
galaxies (UFGs) that deeper studies may reclassify  some of them  as faint halo clusters (FHC), thus supplying   CatClu 
with additional entries. The  present catalogs   form  an interwoven database for future studies. Electronic Table~\ref{tab4} 
(the first 5 lines are shown) compiles  the  kinematical entries. By columns: (1) to (3) U, V and W heliocentric velocities, 
(4) and (5) Galactic coordinates, (6) and (7)  large and small angular diameters, (8) designation(s), (9) type, (10) comments, 
(11)  reference code(s).


\begin{deluxetable*}{rrrrrrrlrlc}
\tabletypesize{\footnotesize}
\tablecaption{245 kinematical Groups and Kinematical Associations\label{tab4}}
\tablehead{
\colhead{U}&\colhead{V}&\colhead{W}&\colhead{$l_G$}&\colhead{$b_G$}&\colhead{d}&\colhead{D}&\colhead{Designations}&\colhead{Type}&\colhead{Comments}&\colhead{Reference} \\
\colhead{(km/s)}&\colhead{(km/s)}&\colhead{(km/s)}&\colhead{$(^o)$}&\colhead{$(^o)$}&\colhead{$(^o)$}&\colhead{$(^o)$}&&&&\colhead{Code}
}
\colnumbers
\startdata
-&-&-&  25.39 & -18.38 & 20   & 10 & A 8 & KGr &78 M III stars; d=92.6 kpc & 3338\\
-7.6 & -27.3 & -14.9 &- &-&360 & 360 &   AB DOR MGr &KGr &  89 stars,d=20 pc &981,697,1730,\\  
     &       &       &  & &    &     &              &    &   rel. to Pleiades?      &1731,1733,1732\\   
-&-&-&-&-& 360& 360&     Acheron Stream	&KGr&         disrupted GC &1744\\
  20 &   -20 &    15&-&-& -&-& AFF 7, ZZC 18& KGr  &     moving group &1722,1723\\         
\hline
\enddata
\end{deluxetable*}

CatKGr deserves an updated catalog owing to many recent studies, especially  halo streams and their occasional 
connection  to former star clusters \citep{BalbGi2018}. CatGal gathers  the recent discoveries of  underluminous 
Milky Way satellites (e.g. \citealt{Drlica-Wagneretal2015}, \citealt{Koposovetal2015s}). Some followup studies have  
reclassified a few as FHCs  (e.g. \citealt{Conn2018}). Another reason to update Table~\ref{tab4} is the growing 
number of compilations and new detections of star clusters and associations with deep surveys (e.g. 1249 young 
clusters in M31 – \citealt{Johnsonetal2016}). CatKGr objects are characterized by heliocentric velocities and/or 
Galactic coordinates, or simply by  designation and reference. If the Sun is spatially located within an object, 
then it becomes  an all-sky object distribution. The halo streams are in general related to dissolved  galaxies  
or globular clusters. 


Electronic Table~\ref{tab5} shows normal, dwarf spheroidal, low luminosity star forming, and UF galaxies.  We adopted 
as  initial catalog \citet{McConnachie2012}'s  with 94 entries. More recently, a number of faint halo galaxies were 
found. The number of galaxies in Table~\ref{tab5} increased LG entries to 144 (53$\%$). Part of the UFGs cannot yet be 
certified as such, or as FHC, since the diagnostic diagram Mv vs. half light radius overlaps them in part. The available 
classifications are sometimes ambiguous.  \citet{Sakamoto2006} could not distinguish the overdensity SDSS J1257+3419 as 
a dwarf galaxy or a globular cluster, but \citet{Belokurovetal2007} classified it as a the galaxy CVn II.  Mass-to-light 
ratios in the range  500-600 \citep{Koposovetal2015} derived from velocity dispersions,  e.g. for Reticulum 2 or Horologium 1, 
indicate dark matter domination, and thus a galaxy classification. Such followup studies with large telescopes can 
spectroscopically ascertain their nature. Table~\ref{tab5}  shows the LG galaxies and references for their clusters and/or  
associations, if any. By columns: (1) and (2) Galactic coordinates, (3) and  (4) J2000.0 equatorial coordinates $\alpha$ 
and $\delta$, (5) and (6)  large and small angular diameters, (7) designation(s), (8) type as normal or dwarf galaxy, (9) 
comments, (10) cluster and association contents and (11)  reference code(s).


\begin{deluxetable*}{rrccrrccccc}
\tabletypesize{\footnotesize}
\tablecaption{Local Group Galaxies and their Clusters and Associations\label{tab5}}
\tablehead{
\colhead{$l_G$}&\colhead{$b_G$}&\colhead{$\alpha$}&\colhead{$\delta$}&\colhead{d}&\colhead{D}&\colhead{Desigations}&\colhead{Type}&\colhead{Comments$^1$}&\colhead{Number}&\colhead{Reference}\\
\colhead{$(^o)$}&\colhead{$(^o)$}&\colhead{(h:m:s)}&\colhead{($^o$:$\arcmin$:$\arcsec$)}&\colhead{$(\arcmin)$}&\colhead{$(\arcmin)$}&&&&\colhead{of SCAOs}&\colhead{Code}
}
\colnumbers
\startdata
5.57 &-14.17 & 18:55:20&  -30:32:43 & 450 & 216&Sagittarius Dwarf,SagDEG &DGAL &dSphE7&incl $\>$6 GCs&1203,1220,1223\\
11.87 &-70.86& 23:26:28& -32:23:20 & 1.8 & 1.8&UKS 2323-326,UGCA 438& DGAL&dIrr,in NGC55  Gr&-& 1220     \\      
18.9& -22.90&19:52:41&-22:04:05& 4 &4 &Sag II,Sagittarius II$^4$ &DGAL & GC?, UFF.&-&3024\\                                                                                                             
21.06& -16.28 &19:29:59& -17:40:41 & 2.9& 2.1&Sagittarius Dwarf Irr.$^2$& DGAL &dIrr IB(s)&-&1220\\
25.34 &-18.40& 19:44:57 &-14:47:21&15&14 &NGC\,6822,IC\,4895$^3$& DGAL&dIrr,inc 6 GCs& 47 young Cls&1220,3342,3353  \\                                      
\hline
\enddata
\tablecomments{ $^1$ The electronic table contains in addition the distance and absolute magnitude.  $^2$ Other designations: ESO\,594-4, SagDIG, UKS\,1927-177. $^3$ Other designations: DDO\,209,Barnard's Galaxy. $^4$ Additional designation: Laevens\,5}
\end{deluxetable*}


\end{document}